\documentclass{article}

\usepackage{spconf,amsmath,graphicx}

\usepackage{hhline}
\usepackage{multirow}
\usepackage[table]{xcolor}
\usepackage{colortbl}
\usepackage{blindtext}
\usepackage{ifthen}
\usepackage{calc}
\usepackage{array}
\usepackage{booktabs}
\usepackage{pifont}
\usepackage{pgfplots}
\usepackage{amsfonts}
\usepackage{caption}
\usepackage{hyperref}
\usepackage{rotating}
\usepackage{setspace}

\usepackage[labelformat=simple]{subcaption}

\usepackage{tikz}
\usepackage{pgfplots}

\pgfplotsset{compat=newest}

\usetikzlibrary{arrows.meta,automata}
\usetikzlibrary{shapes.geometric,shapes.arrows,decorations.pathmorphing}
\usetikzlibrary{matrix,chains,scopes,positioning,arrows,fit}
\usetikzlibrary{patterns}
\usetikzlibrary{chains,backgrounds,calc}

\newcommand{\VEC}[1]{\mathbf{#1}}          
\newcommand{\MAT}[1]{\mathbf{#1}}          

\newcommand{\putindex}[3]{\vtop{\hbox{\hspace{#3} $#1$}
            \hbox{\raise 6mm \hbox{$\scriptscriptstyle #2$}}}}

\newcommand{\gradx}[0]{\vtop{\hbox{\rm grad}
            \hbox{\raise 2.5mm \hbox{\rm \hspace{2mm} \footnotesize x}}}}

\newcommand{\grady}[0]{\vtop{\hbox{\rm grad}
            \hbox{\raise 2.5mm \hbox{\rm \hspace{2mm} \footnotesize y}}}}

\newcommand{\grad}[1]{\vtop{\hbox{\rm grad}
            \hbox{\raise 2.5mm \hbox{#1}}}}

\newcommand{\btb}{     \begin{tabbing}             }
\newcommand{\bte}{     \end{tabbing}               }

\newcommand{\ignore}[1]{}

\newboolean{blind} 
\setboolean{blind}{false}

\def\x{{\mathbf x}}

\newcommand\shorteq{%
	\rule[.35ex]{4pt}{0.4pt}\llap{\rule[.75ex]{4pt}{0.4pt}}}

\newlength{\DepthReference}
\newlength{\HeightReference}
\newlength{\Width}%

\newcommand{\MyColorBox}[2][gray]%
{%
	\settowidth{\Width}{#2}%
	\colorbox{#1}%
	{%
		\raisebox{-\DepthReference}%
		{%
			\parbox[b][\HeightReference+\DepthReference][c]{\Width}{\centering#2}%
		}%
	}%
}

\def\espresso{\texttt{Espresso}\xspace}
\def\fairseq{\texttt{fairseq}\xspace}
\def\pytorch{\texttt{PyTorch}\xspace}

\def\sentencepiece{\texttt{SentencePiece}\xspace}

\title{Relaxed Attention: \\ A Simple Method to Boost Performance of \\ End-to-End Automatic Speech Recognition}
\ifthenelse{\boolean{blind}}{
	\name{XXX, XXX, XXX}
	\address{XXX\\
		XXX\\
		XXX\\
		\fontsize{12}{12}\selectfont {\footnotesize \tt \{X@XX.X, X@XX.X, X@XX.X\}}
		}
}{
	\name{Timo Lohrenz, Patrick Schwarz, Zhengyang Li, Tim Fingscheidt}
	
	\address{Technische Universit\"at Braunschweig\\
		Institute for Communications Technology\\
		Schleinitzstr. 22, 38106 Braunschweig, Germany\\
		\fontsize{12}{12}\selectfont {\footnotesize \tt \{t.lohrenz@tu-bs.de, patrick.schwarz@tu-bs.de, zhengyang.li@tu-bs.de, t.fingscheidt@tu-bs.de\}}
	}
}

\begin{document}

\maketitle

\begin{abstract}\vspace{-1mm}
Recently, attention-based encoder-decoder (AED) models have shown high performance for end-to-end automatic speech recognition (ASR) across several tasks. Addressing overconfidence in such models, in this paper we introduce the concept of relaxed attention, which is a simple gradual injection of a uniform distribution to the encoder-decoder attention weights during training that is easily implemented with two lines of code\footnote{Code contributed to \bf\url{http://github.com/freewym/espresso}}. We investigate the effect of relaxed attention across different AED model architectures and two prominent ASR tasks, Wall Street Journal (WSJ) and Librispeech. We found that transformers trained with relaxed attention outperform the standard baseline models consistently during decoding with external language models. On WSJ, we set a new benchmark for transformer-based end-to-end speech recognition with a word error rate of 3.65\%, outperforming state of the art (4.20\%) by 13.1\% relative, while introducing only a single hyperparameter. 
\end{abstract} 
\vspace{-2mm}

	\vspace{-2mm}
\section{Introduction}
	\vspace{-3mm}
End-to-end automatic speech recognition (ASR) gained a lot of interest in the research community as it makes phonetic modeling obsolete and significantly simplifies the processing pipeline while achieving superior performance compared to hidden Markov model (HMM)-based (hybrid) approaches in many prominent ASR benchmark tasks, especially those that comprise large amounts of data~\cite{chiu2018,Karita2019}. Common end-to-end ASR approaches that directly translate acoustic input sequences into graphemic output sequences are based on connectionist temporal classification (CTC)~\cite{Graves2014}, recurrent neural network transducers (RNN-T)~\cite{Graves2013}, or attention-based encoder-decoder (AED) models~\cite{Bahdanau2015}. The latter approach emerged from neural machine translation and was soon adopted for ASR~\cite{Chorowski2015}. In contrast to early encoder-decoder models that used a fixed-length intermediate representation~\cite{Cho2014}, the attention mechanism uses variable-length attention weight vectors to draw attention to relevant parts in the input sequence, yielding significant improvements for long sentences. Prominent AED model architectures are the RNN-based listen-attend-and-spell (LAS)~\cite{chan2016}, the all-attention-based transformer model~\cite{Vaswani2017,Dong2018c} and, as a variant of the latter, the conformer model~\cite{Gulati2020}.

	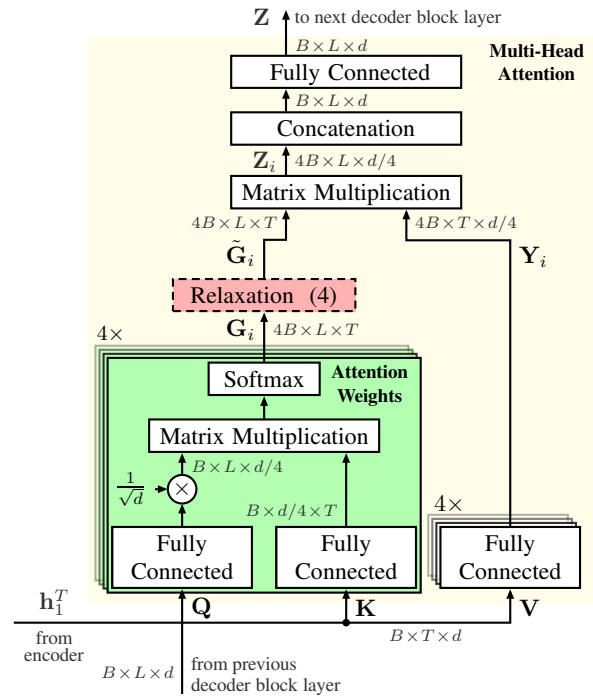
\begin{figure}[t!]
	\centering
	\hspace{0.00cm}
	\begin{tikzpicture}[node distance=0.75cm]
		\tikzset{
			>=stealth',
			tipA/.tip={Triangle[angle=40:4pt]},
			block/.style={rectangle,  draw=black, thick,
				text width=6em, minimum height=0.4cm, text width=6em, text centered,fill=white},
			attn_probs/.style={block, fill=green!30, rounded corners=0cm},
			attn_fusion/.style={block, text width=7em, fill=red!30},
			my_arrow/.style={-|, ->, thick,-tipA},
			dimension/.style={right, text=black!80},
			my_arrow_r/.style={my_arrow, tipA-},
			dot_connect/.style={circle, draw, fill, minimum size=0.08cm, inner sep=0cm},
			bg/.style={fill=yellow!30, rounded corners, draw=black!50, dashed},
			every node/.style={scale=0.9}
		}
		
		{
			\def\x{-8mm}
			
			\def\offsetTwo{0.6*\x}


			\node[block,text width=5.2em] (q_proj_center) {Fully Connected} ;
			\node[draw=none,xshift=-0.2cm,yshift=0.2cm] at (q_proj_center) (q_proj_off){};
			\node[block, text width=5.2em,opacity=1.0] (q_proj) at ($(q_proj_center) !.0! (q_proj_off)$) {Fully Connected} ;

			\node[block,text width=5.2em,right=of q_proj,xshift=-0.5cm] (k_proj_center) {Fully Connected} ;
			\node[draw=none,xshift=-0.2cm,yshift=0.2cm] at (k_proj_center) (k_proj_off){};
			\node[block, text width=5.2em,opacity=1.0] (k_proj) at ($(k_proj_center) !.0! (k_proj_off)$) {Fully Connected} ;

			\node[block, right=of k_proj, text width=5.2em, draw=none, fill=none,xshift=-0.5cm] (value1_center) {Fully Connected} ;
			\node[draw=none,xshift=-0.2cm,yshift=0.2cm] at (value1_center) (value1_off){};
			\node[block, text width=5.2em,opacity=0.3] at ($(value1_center) !.9! (value1_off)$) {Fully Connected} ;
			\node[block, text width=5.2em,opacity=0.5] at ($(value1_center) !.6! (value1_off)$) {Fully Connected} ;
			\node[block, text width=5.2em,opacity=0.8] at ($(value1_center) !.3! (value1_off)$) {Fully Connected} ;
			\node[block, text width=5.2em,opacity=1.0] (v_proj) at ($(value1_center) !.0! (value1_off)$) {Fully Connected} ;
			\node[xshift=-0.9cm,yshift=0.79cm] at ($(value1_center) !.0! (value1_off)$) {$4\times$} ;
			
			\node[draw, circle, fill=white, above=of q_proj, text width=1em, thick,yshift=-0.5cm, minimum size=0.1cm, align=center,inner sep=0mm, outer sep=0mm]  (scale){$\times$};

			\node[block, yshift=2.5cm,text width=9em,yshift=-0.7cm] at ($(q_proj) !.5! (k_proj)$) 		(matmul1)		{Matrix Multiplication\vspace{-0.7mm}};
			\node[block, above=of matmul1,text width=4em,yshift=-0.5cm]  								(softmax)		{Softmax};
			\node[block, above=of softmax,text width=7em,yshift=-0.1cm,densely dashed, fill=red!30]  	(relax)			{Relaxation ~\eqref{eq:relaxAttn}\vspace{-0.7mm}};
			
			\node[block, yshift=5.4cm,text width=9em] (matmul2) at ($(q_proj) !.5! (v_proj)$) 							{Matrix Multiplication\vspace{-0.7mm}} ;
			\node[draw=none,fill=none,xshift=\x,below=of matmul2,yshift=0.5cm,align=center,outer sep=0mm] (tmp_left) {};
			\node[draw=none,fill=none,xshift=-\x,below=of matmul2,yshift=0.5cm,align=center] (tmp_right) {};
			\node[block, above=of matmul2,text width=9em,yshift=-0.4cm] 								(concat) 		{Concatenation} ;
			
			\node[block, above=of concat,text width=9em,yshift=-0.5cm] 									(fc) 		{\vspace{-1mm} Fully Connected} ;
			
			
			\draw[my_arrow,tipA-] (scale.west)  -- ++(-0.17,0cm) node[ pos=1,left] {$\frac{1}{\sqrt{d}}$};
			
			\draw[my_arrow,tipA-] (v_proj.south)  |- ++(-6.6cm,-0.45cm) node[ pos=0,yshift=-2.8mm,right] {$\MAT{V}$} node[align=center,black!80, yshift=-0.3cm,xshift=0.6cm] {\footnotesize \shortstack{\vspace{-0.5mm}from \\\vspace{-1mm}encoder }} node[align=center,black!80, yshift=0.3cm,xshift=0.6cm] {$\VEC{h}_1^T$}  ;
			\draw[my_arrow,tipA-] (k_proj.south)  -- ++(0,-0.45cm) node[ pos=0,yshift=-2.8mm,right] {$\MAT{K}$} node [dot_connect]{} node[dimension, yshift=-0.2cm,xshift=0.5cm] {\scriptsize $B\!\times\!T\!\times\!d$};
			\draw[my_arrow,tipA-] (q_proj.south)  --  
				node[ pos=0,yshift=-2.8mm,right] {$\MAT{Q}$} 
				node[align=left,right,black!80, yshift=0.0cm,xshift=0.00cm,pos=0.8] {\footnotesize \shortstack[l]{\vspace{-0.5mm}from previous \\\vspace{-1mm}decoder block layer}} 
				node[dimension, left, yshift=0.0cm,pos=0.8] {\scriptsize $B\!\times\!L\!\times\!d$}
			++(0,-1.4cm);
			
			\draw[my_arrow] ( q_proj.north)-- ( scale.south) ;
			
			\draw[my_arrow] ( scale.north)-- 
				node[dimension,anchor=west,pos=0.3] {\scriptsize $B\!\times\!L\!\times\!d/4$} 
			([xshift=-1.09cm] matmul1.south) ;
			
			\draw[my_arrow] ( k_proj.north) -- 
				node[dimension,anchor=east,pos=0.2] {\scriptsize $B\!\times\!d/4\!\times\!T$} 
			([xshift=1.09cm] matmul1.south) ;
			
			\draw[my_arrow] ( matmul1.north)-- (softmax.south) ;

			\draw[my_arrow] ( v_proj.north)	|- 
				node[xshift=0.cm,anchor=west,pos=0.5,yshift=-0.248cm]{$\MAT{Y}_i$} 
			(tmp_right)	-| 
				node[dimension,yshift=0.25cm,anchor=west] {\scriptsize $4B\!\times\!T\!\times\!d/4$} 
			([xshift=-\x] matmul2.south) ;
			
			\draw[my_arrow] ( softmax.north)-- node[xshift=0.cm,anchor=east,pos=0.5,yshift=0.1cm]{$\MAT{G}_i$} node[dimension, yshift=0.1cm,anchor=west] {\scriptsize $4B\!\times\!L\!\times\!T$} (relax.south) ;
			
			\draw[my_arrow] ( relax.north)
			|- node[xshift=0.cm,anchor=east,pos=0.5,yshift=-0.2cm]{$\tilde{\MAT{G}}_i$} (tmp_left)
			-| node[dimension, yshift=0.25cm,anchor=east] {\scriptsize $4B\!\times\!L\!\times\!T$} ([xshift=\x] matmul2.south) ;
			
			\draw[my_arrow] ([xshift=\x] matmul2.north) -- 
				node[dimension,pos=0.45]{\scriptsize $4B\!\times\!L\!\times\!d/4$} 
				node[dimension,pos=0.45,left]{$\MAT{Z}_i$} 
			([xshift=\x] concat.south) ;
			
			\draw[my_arrow] ([xshift=\x] concat.north)
			-- node[dimension,pos=0.45]{\scriptsize $B\!\times\!L\!\times\!d$} ([xshift=\x] fc.south) ;
			
			\draw[my_arrow] ([xshift=\x] fc.north)
			-- node[dimension,pos=0.2]{\scriptsize $B\!\times\!L\!\times\!d$} ($([xshift=\x] fc.north) + (0,0.6)$) 
				node[align=left,left,black!80, yshift=0.0cm,xshift=-0.02cm,pos=0.85] {$\MAT{Z}$ } 
				node[align=left,right,black!80, yshift=0.0cm,xshift=0.0cm,pos=0.85] {\footnotesize \shortstack[l]{\vspace{-0.5mm}to next decoder block layer } } ;
			\node[right=of fc,anchor=east, xshift=1.1cm,yshift=0.15cm,align=center] () {\footnotesize \bf  Multi-Head \\[-0.5ex] \footnotesize \bf Attention  };
			
		}

		\begin{pgfonlayer}{background}
			\path (q_proj.west |- fc.north) + (-0.3,0.24) node (g) {};
			\path (v_proj.east |- v_proj.south)+(0.2,-0.2) node (h) {};
			\path[fill=yellow!10,thick]
			(g) rectangle (h);
		\end{pgfonlayer}

		\begin{pgfonlayer}{background}
			\path (q_proj.west |- softmax.north) + (-0.05,0.05) node (g) {};
			\path (k_proj.east |- k_proj.south)+(0.05,-0.05) node (h) {};
			\node [xshift=-0.2cm,yshift=0.2cm](g_off) at (g){};
			\node [xshift=-0.2cm,yshift=0.2cm](h_off) at (h){};
			\path[fill=green!30,draw=black,thick,opacity=0.3] ($(g) !.9! (g_off)$) rectangle ($(h) !.9! (h_off)$);
			\path[fill=green!30,draw=black,thick,opacity=0.5] ($(g) !.6! (g_off)$) rectangle ($(h) !.6! (h_off)$);
			\path[fill=green!30,draw=black,thick,opacity=0.8] ($(g) !.3! (g_off)$) rectangle ($(h) !.3! (h_off)$);
			\path[fill=green!30,draw=black,thick] (g) rectangle (h);
			\node[xshift=0.05cm,yshift=0.35cm] at (g) {\small$4\times$} ;
			\path (softmax.north) -| (k_proj.east) node[pos=0.5,anchor=east, yshift=-0.35cm,align=center] {\footnotesize \bf  Attention \\[-0.5ex] \footnotesize  \bf Weights };
		\end{pgfonlayer}

	\end{tikzpicture}\vspace{-3mm}
	\caption{Encoder-decoder multi-head attention as used in decoder blocks of transformer models (cf.\ Fig.~\ref{fig:decoder_block}) with \textbf{relaxed attention} (red block) \textbf{during training}; $N_\mathrm{h}\!=\!4$.}
	\vspace{-5mm}
	\label{fig:standardMHA}
\end{figure}

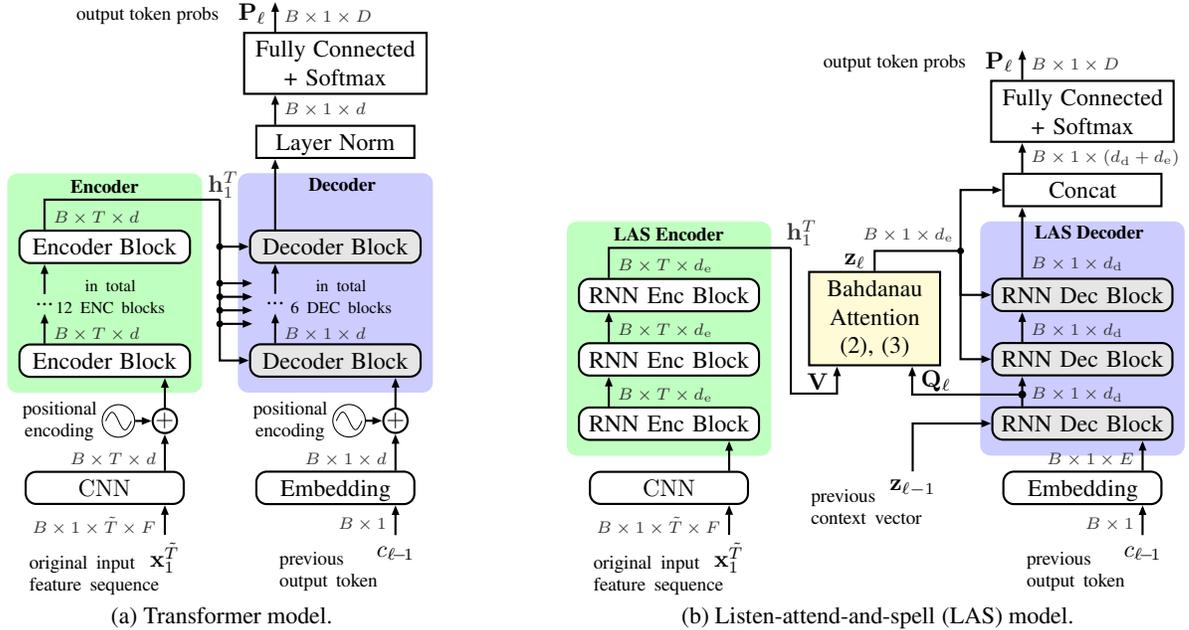
\begin{figure*}[t!]
	\centering

	\begin{subfigure}[b]{0.48\textwidth}
			\centering
		\tikzset{
			tipA/.tip			= {Triangle[angle=40:4pt]},
			punktchain/.style	= {rectangle, rounded corners, draw=black, thick, text width=6.5em, minimum height=0.4cm, text centered,fill=white, on chain=a},
			punkt_mult/.style	= {circle, minimum size=0.1cm,inner sep=0mm, draw=black, text centered,fill=white,thick, on chain},
			arrow/.style		= {draw, thick,-tipA},
			loet/.style 		= {draw,circle,fill=black,inner sep=0.3mm,outer sep=0.0mm},
		}
		
		\begin{tikzpicture}[node distance=0.4cm,scale=1, every node/.style={scale=0.9}]
			{  [start chain=a going above]
				\def\x{-8mm}
				\def\sc{1.11}
				%
				
				\node[on chain=a,outer sep=0mm] (feats) {$\mathbf{x}_1^{\tilde{T}}$};
				\node[punktchain,text width=2.1cm,yshift=-0.0cm,xshift=\x*\sc] (CONV) {$\mathrm{CNN}$};
				\node[punkt_mult,yshift=1mm,xshift=-\x*\sc] (enc_pos_emb) {$\mathbf{+}$};
				\node[draw,circle, left=of enc_pos_emb, xshift=2mm,inner sep= 0mm] (pos_enc) {\tikz \draw[x=0.95mm,y=1.3mm] (0,0) sin (1,1) cos (2,0) sin (3,-1) cos (4,0);};
				\node[left=of pos_enc,draw=none, yshift=0mm, text width=1.1cm,xshift=5mm,align=center] (pos_enc_label){\footnotesize positional \\ \vspace{-1mm} encoding};
				\draw[arrow] ( pos_enc) -- (enc_pos_emb);

				\node[punktchain,draw=black,fill=green!0,xshift=\x*\sc,](Encoder1)      {Encoder Block};
				\node[on chain=a,xshift=\x*\sc](dots){...};
				\node[right=of dots,xshift=-0.75cm, text width=1.8cm,align=center,yshift=1mm](dots_label){\scriptsize in total \\ \vspace{-1mm} 12 ENC blocks};
				
				\node[punktchain,draw=black,fill=green!0,xshift=-\x*\sc](Encoder2)      {Encoder Block};
				\node[left=of feats,draw=none, xshift=1.0cm,yshift=-0.25cm,text width=2.1cm](output_label_label)      {\footnotesize original input \\ \vspace{-1mm}  feature sequence};
				\node[on chain=a,draw=none,xshift=1.7cm](out_node)      {};
				%

				\draw[arrow] ( feats.north) --node[left,text=black!80,pos=0.3] {\scriptsize$B  \times 1 \times \tilde{T} \times F$} ([xshift=-\x] CONV.south);
				\draw[arrow] ( [xshift=-\x]CONV.north) --node[left,text=black!80,pos=0.3] {\scriptsize$B\times T \times d$} (enc_pos_emb.south);
				\draw[arrow] ( enc_pos_emb.north) -- ([xshift=-\x] Encoder1.south);
				\draw[arrow] ([xshift=\x] Encoder1.north) --node[right,text=black!80,pos=0.3] {\scriptsize$B\times T \times d$} ( dots.south);
				\draw[arrow] (dots.north) --node[left,text=black!50] {} ([xshift=\x] Encoder2.south);
				\draw[arrow,-] ([xshift=\x] Encoder2.north) |- node[right,text=black!80,pos=0.2] {\scriptsize$B\times T \times d$}  node[above,text=black!80,pos=1.01,yshift=-0.1cm] {$\MAT{h}_1^T$} (out_node.south);
				
			}
			
			{  [start chain=b going above]
				\def\x{-8mm}
				\def\sc{1.11}
				
				\node[punktchain,on chain=b,right=of CONV,xshift=0.6cm,text width = 2.1cm] (embed) { Embedding\vspace{-0.7mm}};
				\node[draw=none,below=of embed,outer sep=0.5mm,yshift=0.0cm,xshift=-\x*\sc] (input_character) {${c}_{\ell\!-\!1}$};
				\node[punkt_mult,yshift=1mm,on chain=b,xshift=-\x*\sc] (enc_pos_emb) {$\mathbf{+}$};
				\node[draw,circle, left=of enc_pos_emb, xshift=2mm,inner sep= 0mm] (pos_enc) {\tikz \draw[x=0.95mm,y=1.3mm] (0,0) sin (1,1) cos (2,0) sin (3,-1) cos (4,0);};
				\node[left=of pos_enc,draw=none, yshift=0mm, text width=1.1cm,xshift=5mm,align=center] (pos_enc_label){\footnotesize positional \\ \vspace{-1mm} encoding};
				\draw[arrow] ( pos_enc) -- (enc_pos_emb);
				
				\node[punktchain,on chain=b,draw=black,fill=black!10,xshift=\x*\sc](Decoder1)      {Decoder Block};
				\node[on chain=b,xshift=\x*\sc](decoder_dots){...};
				\node[right=of decoder_dots,xshift=-0.75cm, text width=1.8cm,align=center,yshift=1mm](dots2_label){\scriptsize in total \\ \vspace{-1mm} 6 DEC blocks};
				\node[punktchain,on chain=b,draw=black,fill=black!10,xshift=-\x*\sc](Decoder2)      {Decoder Block};
				
				\node[punktchain,on chain=b,draw=black,yshift=0.6cm, text width=2.1cm,,rounded corners=0pt](output_ln)      {Layer Norm\vspace{-0.7mm}};
				\node[punktchain,on chain=b,draw=black,text width=7em,rounded corners=0pt](output_fc)      {Fully Connected + Softmax};

				\node[left=of input_character,draw=none, xshift=1.4cm,yshift=-0.25cm,text width=2.1cm](output_label_label)      {\footnotesize previous \\ \vspace{-1mm} output token};
				
				\draw[arrow] (input_character.north) --node[left,text=black!80,pos=0.36] {\scriptsize$B \times 1 $} ([xshift=-\x]embed.south);
				\draw[arrow] ([xshift=-\x] embed.north) --node[left,text=black!80,pos=0.3] {\scriptsize$B \times 1 \times d$} (enc_pos_emb.south);
				\draw[arrow] ( enc_pos_emb.north) -- ([xshift=-\x] Decoder1.south);
				\draw[arrow] ([xshift=\x] Decoder1.north) --node[right,text=black!80,pos=0.3] {\scriptsize$B \times 1 \times d$} (decoder_dots.south);
				\draw[arrow] ( decoder_dots.north) --node[right,text=black!50] {} ([xshift=\x] Decoder2.south);
				\draw[arrow] ([xshift=\x]  Decoder2.north) --node[right,text=black!80] {} ([xshift=\x] output_ln.south);
				\draw[arrow] ([xshift=\x] output_ln.north) --node[right,text=black!80] {\scriptsize$B \times 1 \times d$} ([xshift=\x] output_fc.south);
				
				\draw[arrow] ([xshift=\x] output_fc.north) --
					node[right,text=black!80] {\scriptsize$B \times 1 \times D$} 
					node[left,draw=none, yshift=-0.23cm,anchor=south east] {$\VEC{P}_\ell$} 
					node[left,draw=none, yshift=-0.23cm,xshift=-.7cm,anchor=south east] {\footnotesize \shortstack{output token probs}}
				++(0,0.4cm);

				\node[loet,left of=Decoder2,xshift=-1.31cm]	(dec_block_2_mag_in)      {};
				\node[loet,left of=decoder_dots,xshift=-0.42cm,yshift=1mm]	(dec_block_mag_dot_in1) {} ;
				\draw[arrow] (dec_block_mag_dot_in1.east) -- ++(4.0mm,0em);
				\node[loet,left of=decoder_dots,xshift=-0.42cm,yshift=3mm]	(dec_block_mag_dot_in2) {} ;
				\draw[arrow] (dec_block_mag_dot_in2.east) -- ++(4.0mm,0em);		
				\node[loet,left of=decoder_dots,xshift=-0.42cm,yshift=-1mm]	(dec_block_mag_dot_in3) {} ;
				\draw[arrow] (dec_block_mag_dot_in3.east) -- ++(4.0mm,0em);	
				\node[loet,left of=decoder_dots,xshift=-0.42cm,yshift=-3mm]	(dec_block_mag_dot_in3) {} ;
				\draw[arrow] (dec_block_mag_dot_in3.east) -- ++(4.0mm,0em);	
				
			}

			\draw[|-,arrow] ( out_node) |- node[right,text=black!50] {} (Decoder1.west);
			\draw[|-,arrow] ( out_node) |- node[right,text=black!50] {} (Decoder2.west);

			\node[above=of Encoder2,anchor=south, xshift=0cm,yshift=-0.02cm,align=right] () {\footnotesize \bf Encoder};
			
			\node[above=of Decoder2,anchor=south, xshift=0.14cm,yshift=0.00cm,align=right] () {\footnotesize\bf Decoder  };
			
			
			\begin{pgfonlayer}{background}
				\path (Encoder2.west |- Encoder2.north)+(-0.15,0.75) node (g) {};
				\path (Encoder1.east |- Encoder1.south)+(0.15,-0.2) node (h) {};
				\path[fill=green!25,rounded corners, draw=black!50,draw=none]
				(g) rectangle (h);
			\end{pgfonlayer}

			\begin{pgfonlayer}{background}
				\path (Decoder2.west |- Decoder2.north)+(-0.15,0.75) node (g) {};
				\path (Decoder1.east |- Decoder1.south)+(0.150,-0.2) node (h) {};
				\path[fill=blue!20,rounded corners, draw=black!50,draw=none]
				(g) rectangle (h);
				
			\end{pgfonlayer}
		\end{tikzpicture}\vspace{-2mm}
		\caption{Transformer model.} 
		\label{fig:transformer}
		
	\end{subfigure} %
	\hspace{0cm}
	\begin{subfigure}[b]{0.48\textwidth}
		
		\centering

		\tikzset{
			tipA/.tip			= {Triangle[angle=40:4pt]},
			punktchain/.style	= {rectangle, rounded corners, draw=black, thick, text width=6.9em, minimum height=0.4cm, text centered,fill=white, on chain=a},
			punkt_mult/.style	= {circle, minimum size=0.1cm,inner sep=0mm, draw=black, text centered,fill=white,thick, on chain},
			arrow/.style		= {draw, thick,-tipA},
			loet/.style 		= {draw,circle,fill=black,inner sep=0.3mm,outer sep=0.0mm},
		}

		\begin{tikzpicture}[node distance=0.4cm, every node/.style={scale=0.9}]
	
			{  [start chain=a going above]
				\def\x{-8mm}
				\def\sc{1.11}
				
				\node[on chain=a,outer sep=0mm] (feats) {$\mathbf{x}_1^{\tilde{T}}$};
				\node[punktchain,text width=2.1cm,yshift=-0.0cm,xshift=\x*\sc] (CONV) {$\mathrm{CNN}$};

				\node[punktchain,draw=black,fill=green!0](Encoder1)      {RNN Enc Block};
				\node[punktchain,draw=black,fill=green!0](Encoder15)	{RNN Enc Block};

				\node[punktchain,draw=black,fill=green!0,](Encoder2)      {RNN Enc Block};
				\node[left=of feats,draw=none, xshift=1.0cm,yshift=-0.25cm,text width=2.1cm](output_label_label)      {\footnotesize original input \\ \vspace{-1mm}  feature sequence};
				\node[on chain=a,draw=none,xshift=1.8cm](out_node)      {};
				
				\draw[arrow] 		( feats.north) --node[left,text=black!80,pos=0.3] {\scriptsize$B  \times 1 \times \tilde{T} \times F$} ([xshift=-\x] CONV.south);
				\draw[arrow] 		([xshift=-\x] CONV.north) -- ([xshift=-\x] Encoder1.south);
				\draw[arrow] 		([xshift=\x] Encoder1.north) --node[right,text=black!80,pos=0.4] {\scriptsize$B\times T \times d_\mathrm{e}$} 	([xshift=\x] Encoder15.south);
				\draw[arrow] 		([xshift=\x] Encoder15.north) --node[right,text=black!80,pos=0.4] {\scriptsize$B\times T \times d_\mathrm{e}$} ([xshift=\x] Encoder2.south);
				\draw[arrow,-] 		([xshift=\x] Encoder2.north) |- node[right,text=black!80,pos=0.2] {\scriptsize$B\times T \times d_\mathrm{e}$} node[above,text=black!80,pos=1.03,yshift=-0.1cm] {$\MAT{h}_1^T$} (out_node.south);
				
			}
			
			{  [start chain=b going above]
				\def\x{-8mm}
				\def\sc{1.11}
				
				\node[punktchain,on chain=b,right=of CONV,xshift=3.3cm,text width = 2.1cm] (embed) { Embedding\vspace{-0.7mm}};
				\node[draw=none,below=of embed,outer sep=0.5mm,yshift=0.0cm,xshift=-\x*\sc] (input_character) {${c}_{\ell\!-\!1}$};
				
				\node[punktchain,on chain=b,draw=black,fill=black!10](Decoder0)      {RNN Dec Block};
				\node[punktchain,on chain=b,draw=black,fill=black!10](Decoder1)      {RNN Dec Block};
				\node[punktchain,on chain=b,draw=black,fill=black!10](Decoder2)      {RNN Dec Block};
				
				\node[punktchain,on chain=b,draw=black,yshift=0.6cm, text width=2.1cm,,rounded corners=0pt]		(output_ln)      		{Concat};
				\node[punktchain,on chain=b,draw=black,text width=7em,rounded corners=0pt]						(output_fc)      		{Fully Connected + Softmax};
				\node[left=of input_character,draw=none, xshift=1.4cm,yshift=-0.25cm,text width=2.1cm]			(output_label_label)    {\footnotesize previous \\ \vspace{-1mm} output token};
				
				\draw[arrow] (input_character.north) --node[left,text=black!80,pos=0.36] {\scriptsize$B \times 1 $} ([xshift=-\x]embed.south);
				\draw[arrow] ([xshift=-\x]embed.north) --node[left,text=black!80,pos=0.36] {\scriptsize$B \times 1 \times E $} ([xshift=-\x]Decoder0.south);
				\draw[arrow] ([xshift=\x] Decoder0.north) --node[right,text=black!80,pos=0.4] {\scriptsize$B \times 1 \times d_\mathrm{d}$} ([xshift=\x] Decoder1.south);
				\draw[arrow] ([xshift=\x] Decoder1.north) --node[right,text=black!80,pos=0.4] {\scriptsize$B \times 1 \times d_\mathrm{d}$} ([xshift=\x] Decoder2.south);
				\draw[arrow] ([xshift=\x] Decoder2.north) --node[right,text=black!80,pos=0.2] {\scriptsize$B \times 1 \times d_\mathrm{d}$} ([xshift=\x] output_ln.south);
				\draw[arrow] ([xshift=\x] output_ln.north) --node[right,text=black!80] {\scriptsize$B \times 1 \times (d_\mathrm{d}+d_\mathrm{e})$} ([xshift=\x] output_fc.south);
				
				\draw[arrow] ([xshift=\x] output_fc.north) --
					node[right,text=black!80] {\scriptsize$B \times 1 \times D$} 
					node[left,draw=none, yshift=-0.23cm,anchor=south east] {$\VEC{P}_\ell$} 
node[left,draw=none, yshift=-0.23cm,xshift=-.7cm,anchor=south east] {\footnotesize \shortstack{output token probs}}
				++(0,0.4cm);
				
				\node[loet,left of=Decoder2,xshift=-1.4cm,yshift=0.65cm]	(dec_block_2_mag_in)      {};

			}
			{
				\def\x{-8mm}
				\def\xx{-5mm}
				\def\sc{1.11}
				
				\node[left=of Decoder1,xshift=-0.3cm,yshift=0.6cm,text width = 1.7cm,draw=black,fill=yellow!20,thick,align=center] (attn) {Bahdanau Attention \\ \eqref{eq:bahdanauScore}, \eqref{eq:bahdanauAtt}};
				
				\node[left=of embed,xshift=-0.42cm,yshift=0.0cm] (prev_context) {$\VEC{z}_{\ell-1}$};
				\node[below=of prev_context,xshift=-0.6cm,yshift=0.8cm,text width = 1.8cm] (prev_context_label) {\footnotesize previous \\ \vspace{-1mm} context vector};
				
				\draw[arrow] (prev_context) |- (Decoder0.west);
				
				\draw[arrow] (out_node) |- ++(0cm,-2.05cm) {} -| node[left, pos=1,yshift=-0.22cm] {$\MAT{V}$}  ([xshift=\xx]attn.south);
				\draw[draw, thick] (attn) |- node[above,text=black!80,pos=0.7] {\scriptsize$B \times 1 \times d_\mathrm{e}$} node[left, pos=0.25] {$\VEC{z}_\ell$} (dec_block_2_mag_in);
				\draw[arrow] (dec_block_2_mag_in) |- (Decoder2);
				\draw[arrow] (dec_block_2_mag_in) |- (Decoder1);
				\draw[arrow] (dec_block_2_mag_in) |- (output_ln);
				
				\draw[thick,arrow] ($([xshift=\x] Decoder0) !.45! ([xshift=\x] Decoder1)$) node[loet]{}  -| node[right, pos=1,yshift=-0.22cm] {$\MAT{Q}_\ell$}  ([xshift=-\xx]attn.south);
				
				
				\node[above=of Encoder2,anchor=south, xshift=0cm,yshift=-0.02cm,align=right] () {\footnotesize \bf LAS Encoder};
				
				\node[above=of Decoder2,anchor=south, xshift=0.14cm,yshift=0.00cm,align=right] () {\footnotesize\bf LAS Decoder  };
			}

			\begin{pgfonlayer}{background}
				\path (Encoder2.west |- Encoder2.north)+(-0.15,0.75) node (g) {};
				\path (Encoder1.east |- Encoder1.south)+(0.15,-0.2) node (h) {};
				\path[fill=green!25,rounded corners, draw=black!50,draw=none]
				(g) rectangle (h);
			\end{pgfonlayer}

			\begin{pgfonlayer}{background}
				\path (Decoder2.west |- Decoder2.north)+(-0.15,0.75) node (g) {};
				\path (Decoder0.east |- Decoder0.south)+(0.150,-0.2) node (h) {};
				\path[fill=blue!20,rounded corners, draw=black!50,draw=none]
				(g) rectangle (h);
				
			\end{pgfonlayer}
		\end{tikzpicture}\vspace{-2mm}
		\caption{Listen-attend-and-spell (LAS) model.}%
		\label{fig:LAS}
		
	\end{subfigure}\vspace{-3mm}
	\caption{\textbf{Standard end-to-end model architectures} used for relaxed attention experiments in this work \textbf{during inference}. Transformer decoder block details are shown in Fig.~\ref{fig:decoder_block}. For faster inference, a batch of $B$ hypotheses are processed in parallel.}
	\vspace{-5mm}
	\label{fig:models}
\end{figure*}

The problem of overconfidence in AED models is demonstrated in~\cite{Li2021}, where utterances with high confidence scores of an LAS model contributed to high word error rates (WER). One reason for such behavior is the use of cross entropy between the predicted output token and the ground truth label as a loss function, as it promotes sparse softmax distributions~\cite{pereyra2017}. This leads to two problems: First, beam search decoding (especially with language models) is less effective as alternatives to a given output token are harder to explore. Second, it is unfavorable for gradient learning as the derivative of the loss function approaches zero when a correct prediction with high confidence is made by the model~\cite{Chorowski2017}. Methods to deal with sharp state probability distributions in (hybrid) ASR (often necessary for stream fusion) are stream weighting~\cite{Receveur2016,Abdelaziz2018,Abdelaziz2015}, limiting~\cite{Lohrenz2017}, and the use of temperature in the softmax function~\cite{Hinton2014,Chorowski2017}. One effective method to deal with overconfidence in \textit{end-to-end ASR}, introduced in~\cite{Szegedy2016}, is label smoothing that blends the one-hot label with a uniform distribution or assigns part of the probability mass to tokens that are neighbors of the labeled token in the target sequence~\cite{Chorowski2017}. Interestingly, label smoothing is also effective against overfitting~\cite{Mueller2020} and thus is commonly used for AED end-to-end ASR besides related regularization methods such as spectral augmentation~\cite{Park2019d}, dropout~\cite{Nitish2014}, multi-task learning with an additional CTC loss~\cite{Kim2017,moriya2020}, and the recently proposed multi-encoder-learning that uses additional encoders only during training~\cite{Lohrenz2021}. Regularization methods applied to the crucial encoder-decoder attention mechanism in AED models were only recently discovered in~\cite{Chen2021}, where CTC predictions in a multi-task learning setup are used to focus the attention in transformer models~\cite{Chen2021} to relevant frames in the encoded input sequence.

In this paper we introduce \textit{relaxed attention}, a simple adjustment to the encoder-decoder attention weights \textit{during training} to reduce overconfidence in AED-based end-to-end speech recognition without adding learnable parameters to the standard model architecture. Different to label smoothing, relaxed attention injects a uniform distribution to the probabilistic attention \textit{weights} (here: not the labels!) to prevent the attention from being overly focused on the encoder input frames. Relaxed attention can easily be implemented in end-to-end ASR toolkits with two lines of code. We investigate the effect of relaxed attention across several attention-based encoder-decoder models, namely the LAS and the transformer model, and across two different tasks (i.e., Wall Street Journal and Librispeech). We also investigate the influence of relaxed attention on the overconfidence problem by ana\-lyzing the AED models with and without integration of external RNN-based language models. 

The paper is structured as follows: Section 2 revises AED models and the attention mechanism to introduce our relaxed attention approach. Section 3 provides details of our conducted experiments, whose results are presented and discussed in Section 4. Section 5 concludes the paper. 
\vspace{-2mm}
\section{Relaxed Attention}	\vspace{-3mm}
\subsection{Attention-Based Encoder-Decoder Models}\vspace{-2mm}
Attention-based encoder-decoder (AED) approaches to end-to-end automatic speech recognition (e.g., the herein used transformer and LAS architectures, shown in Figures~\ref{fig:transformer} and~\ref{fig:LAS}, respectively) comprise an encoder and a decoder network to transform an input feature vector sequence $\VEC{x}_1^{\tilde{T}}$ of dimension $F$ and length $\tilde{T} $ to a sequence of output tokens $c_1^L$ with $c_{\ell}\!\in\!\mathcal{C}\!=\!\{c^{(1)},c^{(2)},\dots,c^{(D)}\}$ being a single output token (i.e., grapheme-based characters or subword units~\cite{Kudo2018}) at output sequence index $\ell\in\{1,\dots,L\}$ from a vocabulary of size $D$. First, the original feature sequence $\VEC{x}_1^{\tilde{T}}$ is commonly preprocessed by several convolutional neural network (CNN) layers to a subsampled representation that is indexed by $t\!\in\!\{1,\dots,T\}$, with $T\!<\!\tilde{T}$. While first approaches towards streaming encoder-decoder models exist~\cite{Kim2019b,Moritz2020}, in this work the encoder network computes a hidden representation $\VEC{h}_1^T\!=\!\mathbf{ENC}(\VEC{x}_1^{\tilde{T}})$ for all $T$ frames that must be available at the start of decoding. For each decoding step (starting at $\ell\!=\!1$), the decoder of the respective model uses the encoded input sequence $\VEC{h}_1^T$ and the previous output token $c_{\ell-1}$ to output a vector $\mathbf{P}_\ell=\mathbf{DEC}(\mathbf{h}_1^T,c_{\ell-1})$ comprising probabilities of all output tokens $c_{\ell}$. These probabilities are then subject to a greedy or beam search algorithm which step-by-step invokes the decoder until some end-of-sentence (EOS) threshold is reached and the final hypothesis is emitted. 

To gather information, which timesteps $t$ in the encoded input sequence are relevant for decoding of the output sequence at step $\ell$, AED models use the attention mechanism that internally computes attention weights containing probabilistic information about relevant input times $t$. 

In the following, we will revise attention types for the two most common AED models that we used in our work. As our proposed relaxed attention is applied only during training, the notations in the following sections hold for the training scenario, where the transformer model is able to train all $L$ output timesteps during decoding in parallel, while the LAS model decodes the output sequence step-by-step. 

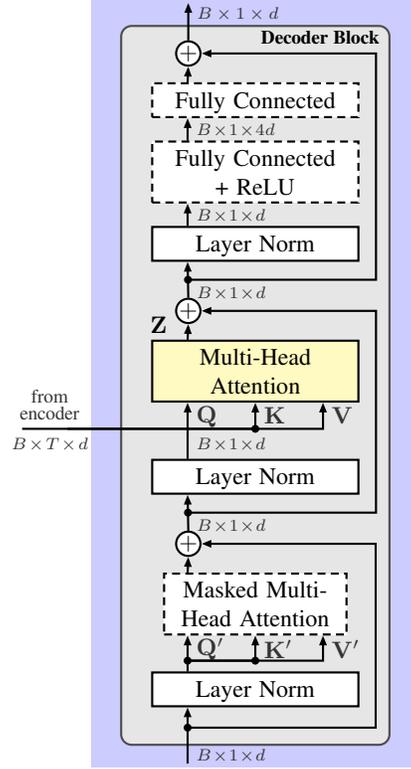
\begin{figure}[t]
	\centering
	\tikzset{%
		tipA/.tip={Triangle[angle=40:4pt]}
	}
	\tikzset{
		>=stealth',
		punktchain/.style={
			rectangle, 
			draw=black, thick,
			text width=8em, 
			minimum height=0.5cm, 
			text centered,fill=white, 
			on chain=encoder_block},
		line/.style={draw, thick, <-},
		arrow/.style={draw, thick,-tipA},
		loet/.style={draw,circle,fill=black,inner sep=0.3mm,outer sep=0.0mm},
		every join/.style={->, thick,shorten >=1pt},
	}
	\hspace{-1.5cm}
	\begin{tikzpicture}[
		node distance=0.75cm,
		mha/.style 		= {draw, fill=yellow!30,,text centered,text width=8em,thick}, 
		scale=1, every node/.style={scale=0.9}
		]

		{  [start chain=encoder_block going above]
			\def\x{-9mm}
			\def\sc{1.1}
			
			\node[punktchain] (ln1) {Layer Norm\vspace{-0.7mm}};
			\node[draw=none,below=of ln1,outer sep=0mm,inner sep=0mm,yshift=0.0cm] (encoder_input) {};
			
			\node[punktchain,densely dashed,yshift=-0.3cm,text width=7em]    (mha) {Masked Multi-Head Attention};
			
			\node[draw,circle,fill=white,on chain=encoder_block,thick,yshift=-0.6cm,xshift=\sc*\x,minimum size=0.1cm,inner sep=0mm]                            (res_in1) {$\mathbf{+}$};
			
			\node[punktchain,xshift=-\sc*\x,yshift=-0.3cm]        (ln_2) {Layer Norm\vspace{-0.7mm}};
			
			\node[punktchain,mha,yshift=0cm] (mhfa) {Multi-Head Attention} ;
			
			\node[draw,circle,fill=white,on chain=encoder_block,thick,yshift=-0.6cm,xshift=\sc*\x,minimum size=0.1cm,inner sep=0mm]                            (res_in2) {$\mathbf{+}$};
			
			\node[punktchain,xshift=-\sc*\x,yshift=-0.3cm] (ln_3) {Layer Norm\vspace{-0.7mm}};
			\node[punktchain,densely dashed,yshift=-0.5cm] (fc_1) {Fully Connected\\ + ReLU};
			\node[punktchain,densely dashed,yshift=-0.5cm] (fc_2) {Fully Connected\vspace{-0.7mm}};
			
			\node[draw,circle,fill=white,on chain=encoder_block,thick,yshift=-0.6cm,xshift=\sc*\x,minimum size=0.1cm,inner sep=0mm] (res_in3) {$\mathbf{+}$};
			
			\draw[arrow] ( [xshift=\x]encoder_input.north) --node[right,text=black!80,yshift=-0.3cm] {\scriptsize$B\!\times\!1\!\times\!d$} ([xshift=\x] ln1.south);
			
			\draw[arrow] ([xshift=\x] ln1.north) --node[right,text=black!80] {} ([xshift=\x]mha.south)node[right,text=black!80,yshift=-0.2cm] {$\MAT{Q}'$}; 
			\draw[arrow] ($ ([xshift=\x] ln1.north) !.3! ([xshift=\x]mha.south) $)node[loet]{} -| node[loet,pos=0.5]{} (mha.south) node[right,text=black!80,yshift=-0.2cm] {$\MAT{K}'$};   
			\draw[arrow] ($ ([xshift=\x] ln1.north) !.3! ([xshift=\x]mha.south) $) -|  ([xshift=-\x]mha.south) node[right,text=black!80,yshift=-0.2cm] {$\MAT{V}'$};  

			\draw[arrow,] ([xshift=\x] mha.north) --node[right,text=black!80,yshift=-0.1cm] {} (res_in1.south);	
			\draw[arrow,] (res_in1.north) --node[right,text=black!80,yshift=-0.2cm] {\scriptsize$B\!\times\!1\!\times\!d$} ([xshift=\x] ln_2.south);

			\draw[arrow] ([xshift=\x]ln_2.north) --  
				node[pos=0.75,right,text=black!80] {$\MAT{Q}$} 
				node[pos=0.25,right,text=black!80] {\scriptsize$B\!\times\!1\!\times\!d$} 
			([xshift=\x] mhfa.south);

			\draw[arrow,tipA-] (mhfa.south) |- 
				node[loet,pos=0.5]{} 
				node[pos=0.25,right,text=black!80] {$\MAT{K}$} 
			([xshift=-2.5cm,yshift=-0.35cm]mhfa.south);
			
			\draw[arrow,tipA-] ([xshift=-\x]mhfa.south) |-  
				node[pos=0.25,right,text=black!80] 																		{$\MAT{V}$} 
				node[left=of mhfa,text=black!80, text width=2cm,  above,pos=0.86, xshift=0cm, yshift=0cm, align=center] 	{\footnotesize \shortstack{\vspace{-0.5mm}from \\\vspace{-1mm} \\ encoder}} 
				node[left=of mhfa,text=black!80, text width=1.7cm,below,pos=0.86, xshift=0cm,yshift=-0.0cm, align=center] 							{\scriptsize$B\!\times\!T\!\times\!d$}
			([xshift=-3.1cm,yshift=-0.35cm]mhfa.south);

			\draw[arrow] ([xshift=\x] mhfa.north) --node[left,text=black] {} node[pos=0.5,left,xshift=-0.18cm,yshift=0.1cm]{$\MAT{Z}$} (res_in2.south);	
			\draw[arrow] (res_in2.north) --node[right,text=black!80,yshift=-0.2cm] {\scriptsize$B\!\times\!1\!\times\!d$} ([xshift=\x] ln_3.south);	
			
			\draw[arrow] ($ ([xshift=\x]encoder_input.south) !.5! ([xshift=\x]ln1) $)node[loet]{} -| ++(2.5,00) |- (res_in1);
			
			\draw[arrow] ($ (res_in1.north) !.5! ([xshift=\x]ln_2.south) $)node[loet]{} -| ++(2.5,00) |-  (res_in2);
			
			\draw[arrow] ($ (res_in2.north) !.5! ([xshift=\x]ln_3.south) $)node[loet]{} -| ++(2.5,00) |- (res_in3);

			\draw[arrow] ([xshift=\x] ln_3.north) --node[right,text=black!80,pos=0.45] {\scriptsize$B\!\times\!1\!\times\!d$} ([xshift=\x] fc_1.south);
			
			\draw[arrow] ([xshift=\x] fc_1.north) --node[right,text=black!80,pos=0.45] {\scriptsize$B\!\times\!1\!\times\!4d$} ([xshift=\x] fc_2.south);
			
			\draw[arrow] ([xshift=\x] fc_2.north) -- (res_in3.south);
			
			\draw[arrow] ( res_in3.north) --node[pos=0.7,right,text=black!80] {\scriptsize$B\times 1 \times d$} +(0,0.5cm);
			
			\node[right=of res_in3,anchor=east, xshift=2.0cm,yshift=0.24cm,align=right] () {\footnotesize \bf Decoder Block };
		}
		
		\begin{pgfonlayer}{background}
			\path (fc_2.west |- fc_2.north)+(-0.8,1.1) node (g) {};
			\path (ln1.east |- ln1.south)+(0.8,-0.8) node (h) {};
			\path[fill=blue!20,]
			(g) rectangle (h);
		\end{pgfonlayer}
		
		\begin{pgfonlayer}{background}
			\path (fc_2.west |- fc_2.north)+(-0.4,0.75) node (g) {};
			\path (ln1.east |- ln1.south)+(0.4,-0.5) node (h) {};
			\path[fill=black!10,rounded corners,draw=black!70, thick]
			(g) rectangle (h);
		\end{pgfonlayer}

	\end{tikzpicture}\vspace{-3mm}
	\caption{Single \textbf{decoder block} as used in the decoder of the transformer model (cf.\ Fig.~\ref{fig:transformer}) \textbf{during inference}, yielding a singleton dimension ($... \!\times\!1\!\times\!...$) in the decoder block tensors, as the decoder is invoked step by step. Dropout layers~\cite{Nitish2014} are in dashed-line boxes. Details of the multi-head attention block (yellow) are shown in Fig.~\ref{fig:standardMHA} for the training case.}%
	\label{fig:decoder_block}
	\vspace{-4mm}
\end{figure}
\vspace{-1mm}
\subsection{Scaled Dot Product Attention (Transformer)}\vspace{-2mm}
The standard scaled dot product multi-head attention (MHA), introduced in~\cite{Vaswani2017}, is the common attention type in several layers of transformer AED models (i.e.,\ each encoder and decoder block cf.\ Figure~\ref{fig:transformer}) to model temporal dependencies without using recurrent layers. While it is implemented as self-attention employed in the \textit{en}coder blocks, here we focus on the encoder-decoder attention in the \textit{de}coder blocks (cf.\ Figure~\ref{fig:decoder_block}) which draws the decoders' attention to relevant parts in the encoded input sequence $\VEC{h}_1^T\!\in\!\mathbb{R}^{T\times d}$. The standard MHA (yellow block) employs multiple attention heads
\vspace{-2mm}\begin{multline}\label{eq:AttHead}
	\mathbf{Z}_i(\MAT{Q},\MAT{K},\MAT{V}) = \\
	\underbrace{\mathbf{softmax}\left(\frac{\MAT{Q}\MAT{W}_i^\mathrm{(Q)}\left(\MAT{K}\MAT{W}_i^\mathrm{(K)}\right)^\mathsf{T}}{\sqrt{d}}\right)}_{\substack{{\text{attention weights}}\\ =\MAT{G}_i(\MAT{Q},\MAT{K})}}\underbrace{\vphantom{ \left(\frac{\left({\MAT{W}^(K)}\right)^T}{\sqrt{d}}\right) }\!\!\!\!\cdot\MAT{V}\MAT{W}_i^\mathrm{(V)}}_{\substack{{\text{value projections}}\\=\MAT{Y}_i(\MAT{V})}}\!\!\!\!\in\mathbb{R}^{L\times\frac{d}{N_\mathrm{h}}}\raisetag{20pt}
\end{multline}\vspace{-2mm}\\
with $\MAT{W}_i^\mathrm{(Q)}, \MAT{W}_i^\mathrm{(K)}$, $\MAT{W}_i^\mathrm{(V)}\!\in\!\mathbb{R}^{d\!\times\! \frac{d}{N_\mathrm{h}}}$ being linear projection weight matrices for the query $\MAT{Q}$, key $\MAT{K}$, and value $\MAT{V}$ inputs, $i\in\mathcal{N}_\mathrm{h}\!=\!\{1\dots N_\mathrm{h}\}$ being the index of the in total $N_\mathrm{h}$ attention heads, and $d$ is the feature vector size being used in most layers of the transformer model. For encoder-decoder attention, key and value inputs stem from the encoder's last layer, yielding $\MAT{K}\shorteq\MAT{V}\shorteq\MAT{h}_1^T$. The entries in each of the $L$ rows of the attention weight matrix $\MAT{G}_i(\MAT{Q},\MAT{K})\!\in\!\mathbb{I}^{L\times T}$, with $\mathbb{I}\!=\![0,1]$, sum up to one and are treated as probabilities that correspond to the relevance of a time frame $t$ to the decoding at step $\ell$. The outputs $\VEC{Z}_i$ of all $N_\mathrm{h}$ separate attention heads are concatenated and subject to a fully connected output layer, yielding the MHA output $\MAT{Z}\in\mathbb{R}^{L\times d}$. Note that for brevity of notation the attention dropout commonly applied to the attention weights in transformer models is not shown in~\eqref{eq:AttHead}. 
 \vspace{-1mm}
\subsection{Bahdanau Attention (LAS)}\vspace{-2mm}
Additive attention, also known as Bahdanau attention, was proposed in~\cite{Bahdanau2015} and is the common attention type for the LAS model, shown here in Figure~\ref{fig:LAS}. One of the main differences of the LAS model compared to the more recent transformer model is the employment of recurrent neural networks (RNNs) in each of the encoder and decoder blocks. Details on the RNN encoder and decoder blocks will be given in Section~\ref{sec:LAS}. Note that for the purpose of clarity, in the following we reuse some notations of the transformer model (e.g.,\ weight matrices), even though these entities depend on their corresponding architecture. Unlike the previously described scaled dot product attention used in transformer models, the LAS model does not incorporate multiple attention heads, but computes a single vectorial context   
\vspace{-1mm}
\begin{equation}\label{eq:bahdanauScore}\vspace{-2mm}
\VEC{z}_\ell=\sum\limits_{t=1}^{T}\mathrm{g}_{t,\ell}\MAT{h}_t \quad\in \mathbb{R}^{1\times d_\mathrm{e} }
\end{equation}
for each timestep $\ell$, with $\mathrm{g}_{t,\ell}$ being an element of the vectorial attention weights $\VEC{g}_\ell\!\in\!\mathbb{I}^{1\times T}$,  and $\MAT{h}_t\in\!\mathbb{R}^{1\times d_\mathrm{e} }$ being a single vector of the encoded input  $\MAT{h}_1^T\!=\!\MAT{V}\!\in\!\mathbb{R}^{T\times d_\mathrm{e} }$ at time index $t$. The attention weights
\vspace{-1mm}
\begin{equation}\label{eq:bahdanauAtt}\vspace{1mm}
\VEC{g}_\ell\,\shorteq\,\mathbf{softmax}\!\!\left(\!\mathbf{v}\!\cdot\!\mathbf{tanh}\!\!\left(\!\VEC{1}\!\cdot\!\mathbf{diag }\!\!\left(\!\VEC{Q}_\ell\!\MAT{W}^{({Q})}\!\!+\!\VEC{b} \!\right)\! +\!\! \VEC{V}\MAT{W}^{({V})}\!\right)^{\!\mathsf{T}} \right)
\end{equation}
are computed utilizing the learnable weights $\MAT{W}^\mathrm{(Q)}\!\in\!\mathbb{R}^{d_\mathrm{d}\!\times\! d_\mathrm{a}}$, $\MAT{W}^\mathrm{(V)}\!\in\!\mathbb{R}^{d_\mathrm{e}\!\times\! d_\mathrm{a}}$, $\MAT{v},\MAT{b}\!\in\!\mathbb{R}^{1\times d_\mathrm{a}}$, with $\VEC{1}$ being a $T\!\times\!d_\mathrm{a}$ matrix with all-ones, $\mathbf{diag}(\MAT{r})$ of $1\!\times\!d_\mathrm{a}$ vector $\MAT{r}$ being its $d_\mathrm{a}\!\times\!d_\mathrm{a}$ diagonal matrix, and $(\,)^\mathsf{T}$ being the transpose. The query input vector $\MAT{Q}_\ell\!\in\!\mathbb{R}^{1\times d_\mathrm{d}}$ stems from the first RNN decoder block of the LAS decoder, and $d_\mathrm{e},d_\mathrm{a},d_\mathrm{d}$ are dimensions of the encoder, attention, and decoder tensors, respectively (cf.\ Fig.~\ref{fig:LAS}).

\subsection{Novel Relaxed Attention}\vspace{-2mm}
According to~\eqref{fig:standardMHA} and~\eqref{eq:bahdanauAtt}, the attention weights for both previously described attention types (i.e., $\MAT{G}_i(\MAT{Q},\MAT{K})\!\in\!\mathbb{I}^{L\times T}$ for the scaled dot product MHA, and $	\VEC{g}_\ell(\MAT{Q}_\ell,\MAT{V})\!\in\!\mathbb{I}^{1\times T}$ for the Bahdanau attention) are of probabilistic nature after the softmax activation function. To prevent overly sharp attention distributions applied in training to the encoded input sequence, our novel relaxed attention weights \textit{for the transformer model} are defined as simple as
\begin{equation}\label{eq:relaxAttn}\vspace{-2mm}
	\tilde{\MAT{G}}_i(\MAT{Q},\MAT{K})=\left[(1-\gamma)\MAT{G}_i(\MAT{Q},\MAT{K}) + \gamma\frac{\VEC{1}}{T}\right],\ \ i\in\mathcal{N}_\mathrm{h},
\end{equation}
 gradually injecting a uniform distribution (with $\VEC{1}$ here being an $L\!\times\! T$ matrix of ones) into the standard attention weights, controlled by a relaxation coefficient $\gamma\in[0,1]$, as shown here in Figure~\ref{fig:standardMHA}. \textit{For the LAS model} with relaxed Bahdanau attention, which we also use for our experiments, the relaxed attention weights for training are defined analogously as 
\vspace{-1mm}\begin{equation}\label{eq:relaxAttn_LAS}\vspace{-1mm}
 \tilde{\VEC{g}}_\ell=(1-\gamma)\VEC{g}_\ell + \gamma\cdot\frac{\VEC{1}}{T},
\end{equation}
with $\VEC{1}$ here being a length $T$ row vector of ones.

\vspace{-1mm}

\section{Experimental Setup}\vspace{-3mm}

\subsection{Databases}	\vspace{-2mm}
We investigate our relaxed attention method on two prominent ASR tasks. First is the 81-hour Wall Street Journal (WSJ) dataset~\cite{paul1992} using the \texttt{dev93} and \texttt{eval92} splits for evaluation. Second is the 980-hour LibriSpeech dataset~\cite{Panayotov2015} with the \texttt{clean} and \texttt{other} conditions of the \texttt{dev} and \texttt{test} datasets. We measure system performance in terms of word error rate $\mathrm{WER}=1-\frac{N-D-I-S}{N}$, as well as w.r.t.\ character error rate (CER) for some experiments, where the number of units $N$, deletions $D$, insertions $I$ and substitutions $S$ are calculated on character-level instead of on word-level as for the WER. All raw speech signals are sampled at 16\,kHz and analyzed with a 25\,ms window and a frame shift of 10\,ms. 
\vspace{-2mm}
\subsection{Acoustic Frontend}\label{sec:frontend}\vspace{-2.5mm}
For all experiments the encoder receives a sequence $\VEC{x}_1^{\tilde{T}}$ of $\tilde{T}$ feature vectors, each of dimension $F\!=\!83$, composed of standard 80-di\-men\-sion\-al filterbank features, extended with 3-dimensional pitch features, both extracted with the \texttt{Kaldi} toolkit~\cite{Povey2011}. The convolutional neural networks (CNNs) at the input layer, shown as CNN block in Figure~\ref{fig:models}, consist of a total of four convolutional layers, each using $3\!\times\!3$ filter kernels. As the second and forth convolutional layer use a stride of 2 in both temporal and frequency direction, the input sequence length is compressed to $T=\tilde{T}/4$. 
\vspace{-2mm}
\subsection{Model Configurations}\vspace{-2mm}
In the following, we will describe all used model architectures and training configurations. All models were trained using the \pytorch-based \espresso and \fairseq tool\-kits~\cite{Paszke2019,ott2019,wang2019e}. For the approaches dubbed \textsf{Baseline}, the model architectures are configured exactly according to the recipes available within the \espresso toolkit. For our new \textsf{Relaxed Attention} approach, we extended the respective baseline models with our simple modification according to~\eqref{eq:relaxAttn} or~\eqref{eq:relaxAttn_LAS}. All models were trained using the Adam optimizer with cross-entropy loss and temporal label smoothing of 0.1 for Librispeech, and 0.05 for WSJ. We follow~\cite{Dong2018c} for tri-stage learning rate scheduling with a maximum learning rate of 0.001. For Librispeech experiments, we also used spectral augmentation~\cite{Park2019d}. No speed perturbation or multi-task learning (as common in~\cite{Karita2019b,Guo2021}) was employed.
\vspace{-2mm}
\subsubsection{Listen-Attend-and-Spell (LAS)}\label{sec:LAS}	\vspace{-2.5mm}
The encoder of the LAS model incorporates three RNN encoder blocks (cf.~\ref{fig:LAS}) each consisting of a dropout layer followed by a single \textit{bidirectional} long-short term memory (LSTM) layer of size $d_\mathrm{e}\!=\!640$. The Bahdanau attention uses an internal attention dimension of $d_\mathrm{a}\!=\!320$. Each of the three RNN decoder blocks first concatenates both inputs before applying a single \textit{unidirectional} LSTM layer with output size $d_\mathrm{d}\!=\!320$. All RNN decoder blocks also employ residual connections and dropout layers before the LSTM layers. The LAS model was trained for 35 epochs and employed a dropout of 0.4.  
\vspace{-2mm}

\newcommand{\f}[1]{\textbf{#1}}
\newcommand{\s}[1]{\underline{#1}}
\renewcommand{\tabcolsep}{0.15cm}
\renewcommand{\arraystretch}{0.95}

\renewcommand{\l}[1]{\small\tt{#1}}

\begin{table}[t!]
	\centering
	
	\begin{tabular}{c c @{\hskip 0.15cm} c c c @{\hskip 0.15cm} c @{\hskip 0.2cm} c @{\hskip 0.15cm} c  }
		\toprule
		\multirow{3}[4]{*}{\shortstack[l]{AED \\ model \\ type}} & \multirow{3}[4]{*}{Approach} & \multirow{3}[4]{*}{\shortstack{\# of \\ AM \\ par's}}   &  \multirow{3}[4]{*}{\shortstack{$\gamma$}} & \multicolumn{2}{c}{\multirow{2}{*}{\tt{dev93}}} & \multicolumn{2}{c}{\multirow{2}{*}{\tt{eval92}}}  \\
		
		& &  &  & & & &   \\
		
		\cmidrule(lr){5-6} \cmidrule(lr){7-8} 
		& 		& 																	& & WER 		& CER 		& WER 		& CER 			\\
		\midrule
		\multirow{8}[10]{*}{\begin{rotate}{90}LAS\end{rotate}} & \sf{Baseline}    	& 17.8M &   0	& 5.70 		& 2.97 		& 4.18  	& 2.11 	\\
		\cmidrule(ll){2-8}
		
		&\multirow{7}{*}{\shortstack[c]{\textsf{Relaxed} \\ \textsf{Attention}}}			& \multirow{7}{*}{17.8M} 		
				& 0.05		& 5.53 		& 2.80 			& 4.02 			& 2.06 	 			\\
		&	&  	& 0.10	 	& \bf5.42 	& 2.75 			& 3.87 			& 1.88 	 			\\
		&	&  	& 0.15	 	& 5.43 		& \bf 2.72 		& \bf 3.77 		& 1.90 	 			\\
		&	&  	& 0.20	 	& 5.59 		& 2.87 			& 3.92 			& 1.96 	 			\\
		&	&  	& 0.25	 	& 5.99 		& 3.14			& 4.02 			& 2.13 	 			\\
		&	&  	& 0.30	 	& 6.00 		& 3.02			& 3.88 			& \bf 1.85	 		\\
		&	&  	& 0.35	 	& 7.10 		& 3.81			& 4.73 			& 2.55	 			\\
		\midrule
		\multirow{9}[17]{*}{\begin{rotate}{90}Transformer\end{rotate}} & \multicolumn{2}{l}{Moriya et al.~\cite{moriya2020}}  &   	& 6.90 		&   		& 4.20  	&   	\\
		\cmidrule(lr){2-8}
		& \sf{Baseline}    	& 16.8M &   0	& 6.69 		& 3.91 		& 4.45  	& 2.46 	\\
		\cmidrule(lr){2-8}
		&\multirow{7}{*}{\shortstack[c]{\textsf{Relaxed} \\ \textsf{Attention}}}			& \multirow{7}{*}{16.8M} 		
				& 0.05	 & 6.41 		& 3.70 			& 4.28 			& 2.34	 			\\
		&	&  	& 0.10	 & 6.54 		& 3.79 			& 4.10 			& 2.26 	 			\\
		&	&  	& 0.15	 & 6.09 		& 3.54 			& 3.96 			& 2.16 	 			\\
		&	&  	& 0.20	 & 6.14 		& 3.45 			& 3.91 			& 2.16 	 			\\
		&	&  	& 0.25	 & 6.02 		& 3.32			& 3.83 			& 2.11 	 			\\
		&	&  	& 0.30	 & 6.32 		& 3.58			& \bf3.65 		& 1.91	 			\\
		&	&  	& 0.35	 & \bf5.80 		& 3.24		& \bf3.65 		& \bf1.85	 		\\
		&	&  	& 0.40	 & 5.83 		& \bf3.19		& 3.74 			& 1.95		 		\\
		\bottomrule
	\end{tabular}\vspace{-2mm}
	\caption{Results on \textbf{WSJ} using \textbf{LAS and transformer} models with various relaxation coefficients $\gamma$ \textbf{with language model}. The number of acoustic model (AM) parameters is shown. Training of each model was repeated 5 times and averaged. Best results for each model type are in \f{bold} font.}
	\label{tab:wsj_lm}
 	\vspace{-3mm}
\end{table}

\subsubsection{Transformer}\label{sec:exp_trans}	\vspace{-2.5mm}
The transformer used in our work follows the standard architecture as introduced in~\cite{Vaswani2017} and is shown in Figures~\ref{fig:transformer},~\ref{fig:decoder_block}, and~\ref{fig:standardMHA}. The encoder uses absolute position embedding on the input that has been preprocessed by the acoustic frontend (cf.\ Sec.~\ref{sec:frontend}) and incorporates 12 encoder blocks, each consisting of multi-head self-attention (MHSA) and pointwise fully connected layers, while the transformer decoder stacks 6 decoder blocks (cf.\ Figure~\ref{fig:decoder_block}). For the WSJ experiments, as well as on the $100$\,h training subset of Librispeech, we employ a smaller model size with $d\!=\!256$, $N_\mathrm{h}\!=\!4$ attention heads, and dropout of $0.1$, while for the larger ($460$ and $960$\,h) datasets of Librispeech we use a large transformer setting with $d\!=\!512$, $N_\mathrm{h}\!=\!8$, and dropout of $0.2$. 

\renewcommand{\tabcolsep}{0.2cm}
\begin{table}[t]
	\centering
	\begin{tabular}{@{\hskip 0.1cm} c @{\hskip 0.15cm} c c c c c @{\hskip 0.3cm}  }
		\toprule
		\multirow{2}[4]{*}{Approach} & \multirow{2}[4]{*}{LM} & \multicolumn{2}{c}{\multirow{1}{*}{\tt{dev93}}} & \multicolumn{2}{c}{\multirow{1}{*}{\tt{eval92}}} \\

		\cmidrule(lr){3-4} \cmidrule(lr){5-6} 
		& 				& WER 			& CER 			& WER 		& CER 			\\
		\midrule
		\multirow{2}[3]{*}{\sf{Baseline}}    		
		& 				& 14.92			& 5.32			& 11.89		& 3.95 \\
		&\checkmark		& 6.69 			& 3.91			& 4.45 		& 2.46 	 		\\
		\midrule
		\multirow{2}[3]{*}{\sf{Relaxed Attention}}			
		&		 		& 16.14			& 5.48 			& 12.73 		& 4.04 	 		\\
		&\checkmark	 	& 5.80 			& 3.24			& 3.65 			& 1.85	 		\\
		\bottomrule
	\end{tabular}	\vspace{-2mm}
	\caption{Results on \textbf{WSJ} \textbf{with and without language model} (LM) using \textbf{transformer} models without (baseline) or with relaxation (coefficient $\gamma\!=\!0.35$). Training of each model was repeated 5 times and averaged.  }
	\label{tab:wsj_nolm}
	\vspace{-4mm}
\end{table}
\vspace{-3mm}
\subsubsection{Conformer}\vspace{-2.5mm}
To further extend our investigations to a larger variety of model architectures, we also employed the recent conformer model~\cite{Gulati2020} for experiments on Librispeech. While using the exact same decoder as the transformer (cf.\ Fig.~\ref{fig:transformer} and Section~\ref{sec:exp_trans}), the conformer model adds a convolutional module after the MHSA in each encoder block, as well as an additional fully connected module before. For the sake of comparability, our implementation uses absolute positional encoding but otherwise follows~\cite{Guo2021} with a total of $12$ encoder blocks and a convolution kernel size of $31$.  
 
 \renewcommand{\tabcolsep}{0.1cm}
 \begin{table*}[t!]
 	\centering
 	
 	\begin{tabular}{l l c c   c @{\hskip 0.1cm} c c @{\hskip 0.1cm} c c c @{\hskip 0.1cm} c c@{\hskip 0.1cm} c   }
 		\toprule
 		\multirow{3}[4]{*}{\shortstack[c]{AED \\ model type}} & \multirow{3}[4]{*}{\shortstack[c]{Training \\ set}} & \multirow{3}[4]{*}{Approach} & \multirow{3}[3]{*}{\shortstack{\# of \\ \textit{acoustic} model \\ parameters}}   &  \multicolumn{4}{c}{\textit{without} LM} & & \multicolumn{4}{c}{\textit{with} LM} \\
 		& & &  &  \multicolumn{2}{c}{\multirow{1}{*}{\tt{dev}}} & \multicolumn{2}{c}{\multirow{1}{*}{\tt{test}}} & & \multicolumn{2}{c}{\multirow{1}{*}{\tt{dev}}} & \multicolumn{2}{c}{\multirow{1}{*}{\tt{test}}} \\
 		\cmidrule(lr){5-6} \cmidrule(lr){7-8} \cmidrule(lr){10-11}  \cmidrule(lr){12-13} 
 		& & 		& 																& \l{clean} 	& \l{other} 	& \l{clean}		& \l{other}			& & \l{clean} 		& \l{other} 		& \l{clean}		& \l{other}		\\
 		\midrule
 		\multirow{6}[5]{*}{Transformer} 
 		& \multirow{2}{*}{$100$ h} 		& \sf{Baseline}    			& 19.31M 		& \bf13.51 		& \bf28.23 		& \bf14.71 		& \bf29.56 			& & 11.18 		& 24.51 		& 12.48  		& 26.65 	\\
 		& 								& \sf{Relaxed Attention}	& 19.31M 		& 14.21 		& 28.81 		& 15.50 		& 30.17 	 		& & \f{9.83}	& \f{21.99}		& \f{10.66}		& \f{23.48}	\\
 		\cmidrule(lr){2-13}
 		& \multirow{2}{*}{$460$ h}  	& \sf{Baseline}				& 69.81M		& \bf4.87 			& \bf13.33 		& \bf5.47 		& 13.53 			& & \bf4.22 		& 11.95 		& 4.95 		& 12.06 	\\
 		& 								& \sf{Relaxed Attention}	& 69.81M 	& 5.12 		 	& 13.42 		& 5.66 			& \bf13.31  			& & 4.50 		& \bf10.60 		& \bf4.85 		& \bf10.75 	\\
 		\cmidrule(lr){2-13}
 		& \multirow{2}{*}{$960$ h}  	& \sf{Baseline}				& 69.81M		& 3.74 			& 8.47 			& 4.14 			& \f{8.48}	 		& & 3.29 		& 7.46 			& 4.02 			& 7.50		\\
 		& 								& \sf{Relaxed Attention}	& 69.81M 		& \f{3.67} 	 	& \f{8.39} 		& \f{4.11} 		& 8.63	 			& & \f{3.12} 	& \f{6.80}		& \f{3.71} 		& \f{7.25} 	\\

 		\midrule
 		\multirow{2}{*}{Conformer} 
 		& \multirow{2}{*}{$960$ h} 		& \sf{Baseline}		 		& 104.7M		& 3.47 			& \bf7.55 			& \bf3.59 		& \bf7.68 	 		 	& & 3.27 		& 6.94 		& \bf3.59 		& 7.17 	\\
 		& 								& \sf{Relaxed Attention}	& 104.7M 		& \bf3.37 			& 7.74 			& 4.06			& 7.85 	 			& & \bf3.16 		& \bf6.59 		& 3.95 		& \bf6.85 	\\
 		\bottomrule
 	\end{tabular}\vspace{-2mm}
 	\caption{WER results on \textbf{Librispeech} using \textbf{transformer and conformer} models; relaxation coefficient $\gamma=0.25$ for the $100$\,h training set, and $\gamma\!=\!0.2$ for all others. Best results for each training set size and model type are in \f{bold} font. }
 	\label{tab:libri_lm}
 	\vspace{-4mm}
 \end{table*}

\begin{table}[t]
	\centering
	\begin{tabular}{@{\hskip 0.05cm} c @{\hskip 0.14cm} c c c c c @{\hskip 0.0cm}  }
		\toprule
		\multirow{2}[4]{*}{Approach} & \multirow{2}[4]{*}{$\gamma$} & \multicolumn{2}{c}{\multirow{1}{*}{\tt{dev}}} & \multicolumn{2}{c}{\multirow{1}{*}{\tt{test}}} \\

		\cmidrule(lr){3-4} \cmidrule(lr){5-6} 
		& 				& \l{clean} 	& \l{other} 	& \l{clean}		& \l{other}				\\
		\midrule
		\sf{Baseline}    		
		& 	0			& 11.18			& 24.51			& 12.48		& 26.65 \\
		\sf{Relaxed Attention}	
		& learned	 	& 10.76			& 24.29 			& 11.46 		& 26.52 	 		\\
				\sf{Relaxed Attention}
		& 0.25				& \f{9.83}	& \f{21.99}		& \f{10.66}		& \f{23.48}	\\
		\bottomrule
	\end{tabular}\vspace{-2mm}
	\caption{WER results of learned relaxation on~\textbf{Librispeech} ($100$\,h training set) using \textbf{transformer} model with LM.  }
	\label{tab:libri_learned}
	\vspace{-3mm}
\end{table}
\vspace{-2mm}
\subsubsection{Tokenization and Language Model}	\vspace{-2mm}
For \textrm{WSJ} experiments, we trained all acoustic models to output tokens on character level, with a total amount of $D\!=\!52$ tokens (including special end-of-sentence and blank symbols). We apply a word-based language model (LM) that is able to output character-level tokens by using the lookahead method from~\cite{hori2018}. The LM is composed of three LSTM layers, each comprising $1200$ neurons totaling in an amount of $112$M parameters. 
For Librispeech we used a total amount of $D\!=\!5000$ subword output tokens generated with \sentencepiece~\cite{Kudo2018}, and employ a token-level LM with four LSTM layers and $800$ neurons each. During decoding, we use shallow fusion~\cite{gulcehre2015,Toshniwal2018} for LM integration according to $\log\VEC{P}_\ell^{\mathrm{final}}\!=\! \log\VEC{P}_\ell  + \lambda\log\VEC{P}_\ell^{\mathrm{LM}}$ with $\lambda$ being the language model weight that we keep fixed to the values from the recipes in \espresso ($\lambda\!=\!0.9$ for WSJ, $\lambda\!=\!0.4$ for Librispeech).

\section{Results and Discussion}\vspace{-3mm}
\subsection{Wall Street Journal (WSJ)}\vspace{-2mm}
Experimental results on the Wall Street Journal task are shown in Table~\ref{tab:wsj_lm}. To achieve statistically profound results, model trainings of both \textsf{Baseline} \textit{and} \textsf{Relaxed Attention} approaches were repeated 5 times (with different seeds for weight initialization) and results were averaged. First, we observe that the WERs for both AED model types (LAS and transformer) are consistently lower with relaxed attention for a wide range of relaxation coefficients. For the LAS model, a WER reduction of 0.28\% absolute (from 5.7\% to 5.42\%) on \texttt{dev93} corresponds to a relative WER reduction of 7.4\% on \texttt{eval92}, when using relaxed attention with $\gamma\!=\!0.1$. For the transformer model, the best average result on \texttt{dev93} is achieved with $\gamma\!=\!0.35$, yielding an average WER of \textbf{5.80}\% on \texttt{dev93} and \textbf{3.65}\% on \texttt{eval92}, which is an 18\% relative improvement on \texttt{eval92} compared to our own \textsf{Baseline} model (4.45\%), and exceeds the current WSJ transformer state of the art by Moriya et al.~\cite{moriya2020} (4.20\%) by 13.1\% relative, without adding any model complexity. 

Interestingly, we note that the WERs on the \texttt{eval92} set are consistently decreasing with increasing relaxation (until $\gamma\!=\!0.35$) and the single-best result for the transformer model (before averaging, not shown in Table~\ref{tab:wsj_lm}) even reaches 5.65\% on \texttt{dev93}, with a benchmark WER of \textbf{3.19\%} on \texttt{eval92}.

In a small ablation study shown in Table~\ref{tab:wsj_nolm}, we investigate the behavior of both \textsf{Baseline} and \textsf{Relaxed Attention} approaches \textit{with} and \textit{without} language model (LM).\footnote{In supplementary experiments we smoothed attention weights with a tempered softmax and obtained results inferior to \textsf{Baseline} performance.} For the optimal transformer relaxation coefficient $\gamma\!=\!0.35$ that has been found before under use of an LM, relaxed attention training performs suboptimal w/o LM, while with LM we obtain the benchmark results from Table~\ref{tab:wsj_lm}. This indicates that even though relaxed attention does not improve performance w/o LM on WSJ, it helps decreasing overconfidence and makes the model perfectly suitable for language model integration.

We additionally performed a further analysis of entropy in the \textit{transformer} MHA weights w/o LM. During training, by application of~\eqref{eq:relaxAttn}, the relaxed attention weights $\tilde{\MAT{G}}_i$ have higher entropy compared to $\MAT{G}_i$, as expected. During \textit{inference} on $\texttt{dev93}$, the attention weights of the best \textsf{Relaxed Attention} model ($\gamma\!=\!0.35$) yield a 4\% higher entropy as compared to the \textsf{Baseline} model, thereby confidence is decreased even without relaxation~\eqref{eq:relaxAttn} in inference, giving important degrees of freedom to the LM. 
  
\vspace{-2mm}
\subsection{Librispeech}\vspace{-2.5mm}
We choose Librispeech to validate our relaxed attention approach on a large dataset and also evaluate performance on increasing training set sizes in Table~\ref{tab:libri_lm}. We report on transformer-based models as they yield superior performance compared to LAS models on Librispeech (e.g.,\ in~\cite{Karita2019,Zeyer2019}). We also use a re-simulated conformer model, which holds the benchmark WERs of 1.9\%/3.9\% in~\cite{Gulati2020} on the \texttt{clean} / \texttt{other} portions of the \texttt{test} set. All re-simulated models are compared without and with relaxed attention.

In our experiments we note that even \textit{without} LM, relaxed attention helps in some cases on the larger training sets for both model types, while showing similar behavior as on WSJ on the similar-sized $100$\,h training set. \textit{With} LM on the \texttt{dev} set, in 7 out of 8 cases relaxed attention leads to improvements over all training set sizes and models, with particularly consistent improvements in the \texttt{other} condition. On the \texttt{test} set with LM, our relaxed attention for \textit{transformer} models exceeds \textsf{Baseline} performance in all conditions and all training set sizes, while for the conformer model only an improvement in the \texttt{other} condition is achieved ($\gamma$ hasn't been optimized for the conformer). Using a standard transformer model trained with the entire $960$\,h training set, relaxed attention achieves a relative improvement of 4.9\% averaged across both \texttt{test} set conditions with LM (5.76\% vs.\ 5.48\% absolute WER).

In Table~\ref{tab:libri_learned}, we learned the relaxation coefficient $\gamma$ during training and observe that performance of the learned \textsf{Relaxed Attention} is close to---but yet still lower---than the \textsf{Baseline} approach. This is expected, as relaxed attention (similar to other generalization techniques, e.g., dropout) harms the training loss and thus the learned $\gamma$ values in each decoder block converged towards small values in a range of $[0, 0.03]$ during training in our experiments. We conclude, however, that $\gamma$ should not be learned but manually set to put stress on the network to still learn under relaxed attention. 
\vspace{-2mm}
\section{Conclusion}\vspace{-3mm}
In this work we introduced relaxed attention for end-to-end ASR, a simple method that smoothes attention weights in attention-based encoder-decoder models during training to decrease overconfidence of these models. Across a variety of encoder-decoder models, we observe performance gains when our method is used in combination with external language models. Particularly on the WSJ task, transformer models trained \textit{with} relaxed attention reduce the average word error rate by 13.1\% relative compared to state of the art, setting a new benchmark of 3.65\% WER on WSJ for transformer-based automatic speech recognition without adding any model complexity in inference.

\ifthenelse{\boolean{blind}}{
	}{
	{\begin{center}
			{\bf ACKNOWLEDGMENTS\vspace{-.5em}\par}
	\end{center}}\par
	\vspace{-1mm}
	\noindent The research leading to these results has received funding by the Deutsche Forschungsgemeinschaft (DFG, German Research Foundation) for project number 414091002. 
	}

\clearpage

\bibliography{ifn_spaml_bibliography}
\bibliographystyle{IEEEbib}

\end{document}